# The investigation of relativistic electron electromagnetic field features during interaction with matter.

G.A. Naumenko[1], A.P. Potylitsyn[1], L.G. Sukhikh[1], Yu.A. Popov[1]

[1]Nuclear physics institute, Tomsk Polytechnic university, Lenina str. 2a, Tomsk, 634050, Russia.

[2]Institut de Physique Nucl.eaire de Lyon, Université de Lyon,CNRS- IN2P3 and Université Lyon 1, 69622 Villeurbanne, France

The features of electromagnetic field of relativistic electrons passing through a hole in an absorbing screen as a function of the distance from the screen in range of radiation formation length were investigated. The analysis of obtained results allows approving the existence of an unstable state of electron with a particularly deprived its coulomb field, which turns into a stable state of usual electron at a distance of a radiation formation length.

**Introduction.**

If a relativistic charged particle with the Lorents-factor $\gamma$ interacts with a scattering center, the mixed state of a total electromagnetic field (radiation field + particle field) downstream to the interaction point may be observed. At the distance $l_f = \gamma^2 \lambda$ ($\lambda$ is the wavelength of the field Fourier harmonic) this field evolutes to a stable state of radiation field and usual charged particle field. In principle the exact solution of the Maxwell equations may be used for description of this process. However these solutions are often very complicated, or sometimes are difficult at the stage of a problem definition. Therefore phenomenological concepts like *equivalent photons* and *surface current* viewpoint are useful for an intuitive understanding of the main features. . The last one was clearly shown by B.M. Bolotovskiy in [1] for example of forward Diffraction Radiation (DR) of relativistic electron moving near a conductive semi-plane (Fig.1). The traveling Coulomb field (represented by ellipses centered on the successive positions of the particle) induces current in the half-plane screen, which in turn emit DR, represented by small pieces of ellipses. The radiation field is such that close to the screen it kills part of the particle field. The interference gradually disappears (positions 3, 4, 5 of the figure) due to the different velocities, $v \simeq 1 - \gamma^{-2}/2$ and $c = 1$ (in our units), of the Coulomb and radiation fields. These fields get out of phase after a time $t_f \sim \lambda/(c-v) \sim l_f$.

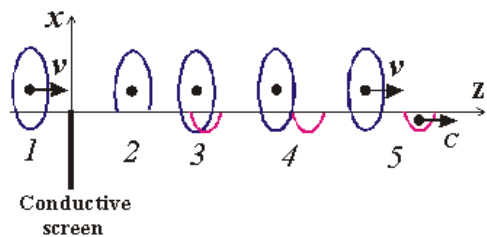

Figure 1 Illustration of the radiation formation length effect by B.M. BolotovskiI in [1]

"Method of images" is also used to calculate a forward and backward TR and DR [2,3]. This method comes from electrostatics and is based on a surface charge and surface current concept.

Another point of view is the equivalent photons method. The Coulomb field is considered as a beam of quasi-real photon. For ultra-relativistic electrons the properties of these photons are very close to the properties of real photons. Namely, in wavelength rang from optics to millimeter wavelength, they are reflected from a mirror, absorbed in absorber and they don't induce a surface current on a downstream surface of a thick conductive target. There is a region downstream to a conductive or absorbing screen where the Coulomb field is partly missing. In terms of paper [4] this effect is named "shadow effect", and the term "half-naked electron" has been introduced in [5,6] to describe the similar effect in the framework of quantum electrodynamics for an electron scattered at a large angle. In both these interpretations the Coulomb field is gradually "repaired" during the formation zone length $l_f \sim \gamma^2 \lambda$.

In [7] a shadowing of an electron Coulomb field by the conductive and absorbing semi-plane in macroscopic mode was investigated. It is remarkable, that no principal difference was found in experiment, whether we use a conductive or absorbing screen for shadowing. Moreover, in [8] was shown experimentally that the electron field does not induce a surface current on a downstream conductive target surface. Therefore we may expect that downstream to the absorbing screen A (see Fig. 2) the electromagnetic field is defined by the evolution of the electron field.

Presented experiment is devoted to the measurement of a total electromagnetic field evolution inside a formation zone when an electron passes through a hole in an absorbing screen A (Fig. 2).

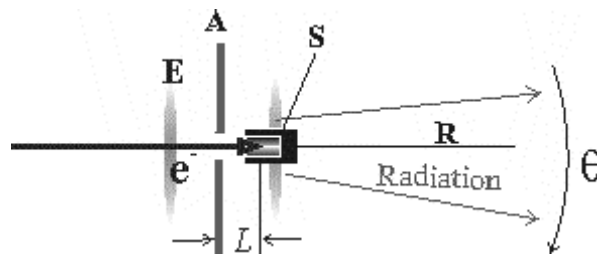

Fig. 2. A schematic of a possible measurement of an electron electromagnetic field evolution. **E** is a shape of an electron electromagnetic field, **A** is the absorbing screen, **S** is the beam dump.

For this purpose we stop the electron beam by the beam-dump S, which provides the full absorption of relativistic electrons. In frame of equivalent photons viewpoint, a photon shape of electron field continues evolution, being transformed into real photons. Measuring this radiation in far field zone we obtain information about the state of electromagnetic field (electron field + radiation) at the distance L from the screen.

**Experiment.**

The experiment was performed on the extracted electron beam of the microtron of the Tomsk Nuclear Physics Institute (Russia). The beam is extracted from the vacuum chamber through a 20 $\mu$m thick beryllium foil. The beam parameters are listed in table 1.

**Table 1.** Electron beam parameters.

| Electron energy | 6.1 MeV ($\gamma = 12$) | Bunch period | 380 psec |
| --- | --- | --- | --- |
| Train duration | $\tau \approx 4$ $\mu$sec | Bunch population | $N_e$=6·10$^8$ |
| Bunches in a train | $n_b \approx 1.6 \cdot 10^4$ | Bunch length | $\sigma \approx$ 1.3~1.6mm |

The window caused a beam divergence ($\simeq 0.08$ radian). For listed bunch length and population, the electron field and radiation with a wavelength $\lambda > 8mm$ are coherent and radiation intensity is enhanced by $10^8$ times. This allows us to measure the radiation using a room-temperature detector. For the radiation measurements we used the detector DP20M, with parameters described in [9]. The detector efficiency in the wavelength region $\lambda$=3~16 mm is estimated to be constant to a ±15% accuracy. The detector sensitivity is 0.3 V/mWatt. A wave-guide with a cutoff $\lambda_{cut}$=17 mm was used to cut the long-wave background of the accelerator RF system. The high frequency limit of the wavelength interval is defined by the bunch form-factor. This limit ($\lambda_{min}$=9 mm) was measured using discrete wave filters [11] and a grating spectrometer.

To exclude the prewave zone effect (see [10]) a parabolic telescope was used to investigate the radiation angular distribution (see Fig. 3). This method was suggested and tested in [12] and provides the same angular distribution as in the far field zone ($R \gg \gamma^2\lambda$).

The used absorber provides an absorption in investigated wavelength range by 60 decibel, without radiation reflection. The Faraday cup was used for monitoring possible electrons skipping the beam-dump. To minimize a contribution of these electrons in measured characteristics due to a beam divergence, the beam-dump S was placed at a fixed position and the variation of distance $L$ was provided by the changing of screen position. This has limited a maximal value of distance $L$ by 200 mm.

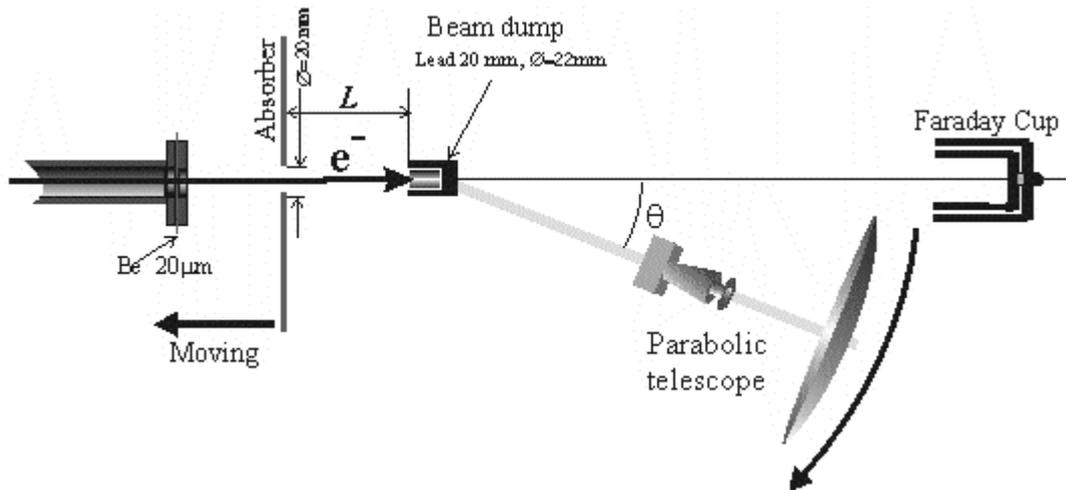

Fig. 3. Scheme of experiment.

The radiation angular distribution was measured with step 1° for different values of distance $L$ from 0 to 200 mm with step 20 mm. In Fig. 4 are shown the samples of measured angular distribution for different values of distance $L$.

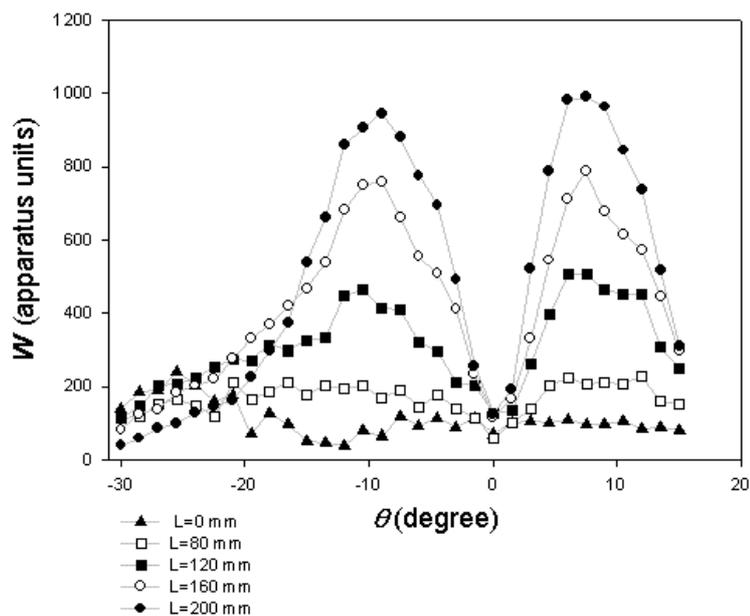

Fig. 4. Samples of the measured angular distribution of radiation intensity for different distances $L$ between screen and beam dump.

The full smoothed dependence is shown in Fig. 5.

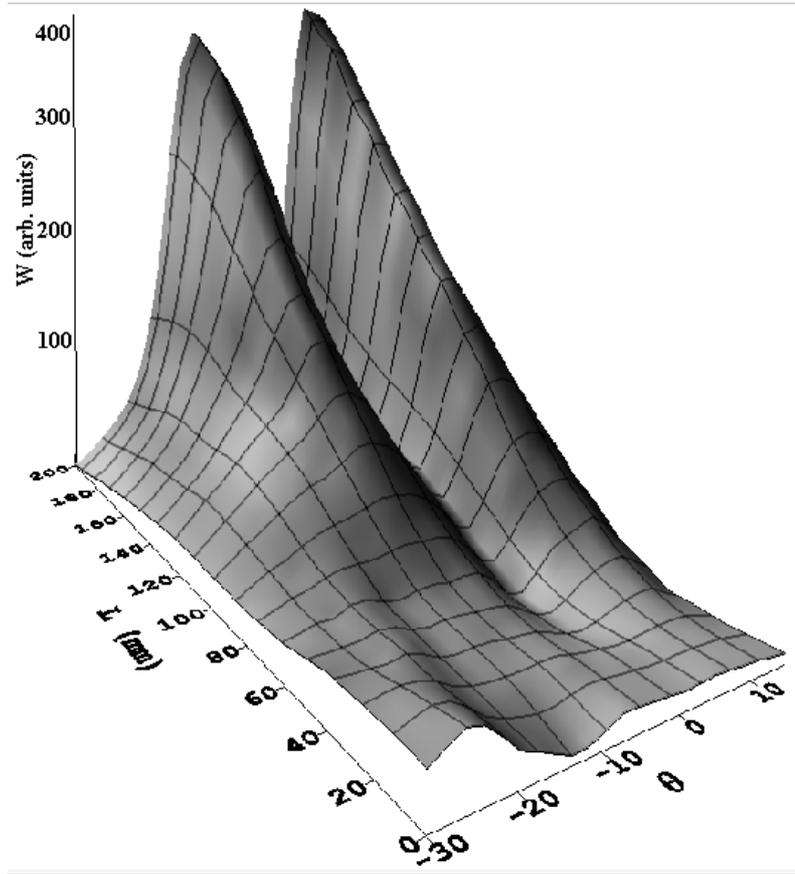

Fig. 5. The smoothed two-dimensional measured dependence of radiation intensity on the observation angle θ and distance L.

**Discussion.**

The measured dependence shown in Fig. 5, presents the properties of an electromagnetic field, including the intrinsic Coulomb electron field and radiation field at a distance $L$ from the absorbing screen. We see that the electromagnetic field is built up when the distance L increases from zero to 200 mm. The absence of a surface current on the screen (as it was shown above) doesn't allow us to consider this process as interference between a radiation from the screen surface and an intrinsic Coulomb electron field, like in [1]. Moreover, since in our experimental conditions the electron moving is uniform, we can't explain the experimental results as an interference of an intrinsic Coulomb electron field with a radiation of a suddenly accelerated electron in screen plane, like in Tamm problem [13]. Besides that, the absorption of electrons in beam-dump may be not interpreted as a transition radiation source due to a negative acceleration of a suddenly stopping electron, because the electrons are stopped inside the beam-dump. Taking into account the above analysis, we may conclude that the dependence shown in Fig. 5, presents a transition of an electron from state of half-naked electron to a state of a usual electron, similar the E.L. Feinberg interpretation in [5]. The used experimental strategy does allow us separate the intrinsic Coulomb electron field and possible radiation attendant on this transition inside the interval $0 < L < 200 mm$. Unfortunately the experimental conditions don't allow us prolongation of these measurements to $L = \gamma^2 \lambda$ (in our case $\gamma^2 \lambda \approx 1.5 m$) for registration of the stable electron state and this radiation.

**Acknowledgment**

This work was partly supported by the warrant-order 1.226.08 of the Ministry of Education and Science of the Russian Federation and by the Federal agency for science and innovation, contract 02.740.11.0245

**References**
[1] BolotovskiI B N. Preprints of Lebedev Institute of Physics, Soviet Academy of Sciences,Vol **140** p. 95
[2] V.L. Ginzburg, V.N.Tsitovich. Transition radiation and transition scattering. Moskow, Science,. 1983
[3] Bolotovskii B M, Serov A V "Features of the transition radiation field" Phys. Usp. **52** 487 (2009)
[4] X. Artru, R. Chehab, K. Honkavaara, A. Variola, Nucl. Inst. Meth. B145 (1998) 160.
[5] Feĭnberg E L. SOV PHYS USPEKHI, 1979, **22** (6), 479-479
[6] Shul'ga N F and Syshchenko V V. Journal Physics of Atomic Nuclei, **63**, 11, (2000), 2018
[7] Naumenko G.A., Potylitsin A.P., Sukhikh L.G., Popov Yu.A., Shevelev M.V. Macroscopic Effect of the Shadow of the Electromagnetic Field of Relativistic Electrons. JETP Letters, 2009. - т.90 - № 2. - с. 96-101 (21705367)
[8] G.A. Naumenko, A.P. Potylitsyn, L.G. Sukhikh, Yu.A. Popov. The experimental study of the surface current excitation by a relativistic electron electromagnetic field. P. 543 in book "Charged and Neutral Particles Channeling Phenomena - Channeling 2008", Proceedings of the 51st Workshop of the INFN Eloisatron Project, S.B. Dabagov and L. Palumbo, Eds., World Scientific, 2010. See also http://arxiv.org/pdf/0901.2630v1.
[9] Kalinin B N, Naumenko G A, Potylitsyn A P et al. *JETP Letters*, 2006, Vol. **84**, No. 3, pp. 110–114.
[10] Hanke K. *CLIC note* 298, 19. 04.1996
[11] Verzilov V A. *Phys. Lett*. A **273** (2000) 135.
[12] Kalinin B N, Naumenko G A, Potylitsyn A P et al. *JETP Letters*, **84**, 3, (2006), p. 110.
[13] Tamm I. E.: J. Phys. USSR*1* (1939) 439.